\documentclass[twocolumn,showpacs,preprintnumbers,superscriptaddress,
               amsmath,amssymb,prd]{revtex4}

\usepackage[dvips]{graphicx}
\usepackage{dcolumn}
\usepackage{bm}

\def\bfm#1{\mbox{\boldmath $#1$}}
\def\bfsm#1{\mathstrut\mbox{\scriptsize{\boldmath $#1$}}\mathstrut}

\newcommand{\ga}{\gamma}

\newcommand{\de}{\delta}
\newcommand{\De}{\Delta}

\newcommand{\la}{\lambda}
\newcommand{\La}{\Lambda}

\newcommand{\Om}{\Omega}
\newcommand{\p}{\partial}
\newcommand{\lt}{\langle} 
\newcommand{\gt}{\rangle}  

\newcommand{\Eqn}[1]{Eq.~(\ref{#1})}  
\newcommand{\ba}{\begin{array}}
\newcommand{\ea}{\end{array}}
\newcommand{\bea}{\begin{eqnarray}}
\newcommand{\eea}{\end{eqnarray}}
\newcommand{\bal}{\begin{align}}  
\newcommand{\eal}{\end{align}}
\newcommand{\bi}{\begin{itemize}}  
\newcommand{\ei}{\end{itemize}}
\newcommand{\ben}{\begin{enumerate}}  
\newcommand{\een}{\end{enumerate}}

\newcommand\hide[1]{}

\newcommand{\Tr}{\mbox{Tr}}

\newcommand{\ie}{{i.e.}}

\newcommand{\nn}{\nonumber \\}

\newcommand{\feyn}[1]{
  \setbox0=\hbox{\ensuremath{#1}}
  \hbox to\wd0{\hbox to0pt{\hbox to\wd0{\hss/\hss}\hss}\box0}}

\newcommand{\MeV}{\,{\rm MeV}} 
\newcommand{\GeV}{\,{\rm GeV}}

\newcommand{\Qtilde}{{\tilde Q}}
\newcommand{\mue}{\mu_{e}}
\newcommand{\Ms}{M_s}
\newcommand{\btem}{\bibitem}

\begin{document}


\title{How do chiral condensates affect color superconducting
 quark matter \\
under charge neutrality constraints?}

\author{Hiroaki Abuki}
\email[E-mail:~]{abuki@yukawa.kyoto-u.ac.jp}
\affiliation{Yukawa Institute for Theoretical Physics, Kyoto University, 
Kyoto 606-8502, Japan}
\author{Masakiyo Kitazawa} 
\email[E-mail:~]{masky@yukawa.kyoto-u.ac.jp}
\affiliation{Yukawa Institute for Theoretical Physics, Kyoto University, 
Kyoto 606-8502, Japan}
\affiliation{Department of Physics, Kyoto University, Kyoto 606-8502,
Japan}
\author{Teiji Kunihiro}
\email[E-mail:~]{kunihiro@yukawa.kyoto-u.ac.jp}
\affiliation{Yukawa Institute for Theoretical Physics, Kyoto University, 
Kyoto 606-8502, Japan}

\date{\today}

\begin{abstract}
We investigate the effects of the dynamical formation of the chiral
 condensates on color superconducting phases under the electric and
 color neutrality constraints at vanishing temperature.
We shall show that the phase appearing next to the color-flavor-locked
 (CFL) phase down in density depends on the strength  of the diquark
 coupling.
In particular, the gapless CFL (gCFL) phase is realized only in a weak
 coupling regime.
We give a qualitative argument on why the gCFL phase in the weak
 coupling region is replaced by some other phases in the strong
 coupling, once the competition between dynamical chiral symmetry
 breaking and the Cooper pair formation is taken
 into account.
\end{abstract}

\maketitle

On the basis of the asymptotic-free nature of QCD and the attraction
 between quarks due to gluon exchanges, we now believe that the ground
 state of the quark matter composed of u, d and s quarks at extremely
 high densities is a special type of color superconducting phases
 \cite{BL84,II95};
 that is the color-flavor locked (CFL) phase where all the quarks
 equally participate in pairing \cite{reviews,ARW98}.

In reality, nature may not, however, allow such an extremely
 high-density matter to exist, even in the core of neutron stars and in
 possible quark stars.
In such systems at relatively low density corresponding to the quark
 chemical potential of, say, $500\MeV$, the following two ingredients
 become important for the fate of the CFL phase and determining the
 pattern of color superconductivities \cite{IB02,RW01,AR02}:
Firstly, one can not neglect the effect of the strange quark mass $\Ms$
 which ranges from around $100\MeV$ to $500\MeV$ depending on the quark
 density.
Secondly, the constraints of the color and electric neutrality 
 must be satisfied as well as $\beta$-equilibrium under the weak
 interaction.
The former causes Fermi-momentum mismatch \cite{ARK00,SW99,abuki02}, and
 the latter pulls up or down the Fermi momentum of each species of
 quarks \cite{RW01,AR02};
as the density goes lower, the symmetric CFL pairing would cease to be
 the ground state at some critical density, and some phases other than
 the CFL phase would appear.

One of the recent findings of such novel pairing patterns is the
 gapless CFL (gCFL) phase \cite{AKR04a,AKR04b}, which is a non-BCS state
 having some quarks with gapless dispersions despite  the same symmetry
 breaking pattern as the CFL phase. 
Historically, a possible realization of the  stable gapless
 state was first discussed for the two-flavor color superconducting
 phase \cite{HS02}:
It was shown that the local charge neutrality gives a so strong
 constraint that such an exotic state, called the g2SC phase, exists
 stably;
this is in contrast with the case of the electronic superconductivity
 in metals \cite{Sarma}, where the possible gapless state is unstable
 against the spatial separation into the Pauli-paramagnetic and
 superconducting phases because of the absence of a long-range force
 mediated by gauge fields.  
The gCFL phase is the three flavor analogue of the g2SC phase.
Successive detailed studies have revealed a rich phase structure of
 superconducting quark matter at zero and nonzero temperatures
 \cite{RSR04,FKR04}. 
It should be also noted that the possible existence of the gCFL phase
 in a compact star may lead to astronomically interesting consequences
 because of the existence of the gapless quarks \cite{AKR04c}.
Thus, examining the robustness of the gapless phases
as well as exploring their
 nature has become one of the central subjects in the study of QCD
 matter in extreme conditions.
In fact, it has been recently indicated \cite{Unstable} that 
the gluons in the gapless phases acquire an imaginary Meissner mass,
which may signal an instability of the system to a yet unknown state.

In this Letter, we investigate how the superconducting orders including
 the gapless phases are affected by the incorporation of the dynamical
 chiral condensation.
This incorporation should be important when the color superconductivity
 in a compact star is considered, where a change of the chiral condensate
 $\lt\bar{q}q\gt$ is also expected.
In fact, the dynamically generated chiral condensate may greatly affect
 the stability of the gapped superconducting phases, leading to a quite
 novel phase structure:
(i)~The interplay between the chiral and diquark condensations
 makes the quark masses depend on the realized phases
 \cite{abuki00,kkkn02,SRP02}, and hence some phases in turn become
 disfavored or favored.
(ii)~ The gapless system might become unstable against the
 phase separation into the phases with a different chiral condensate
 $\lt\bar{q}q\gt$ since the scalar condensate has no gauged charges;
 recall the fate of the possible gapless state in the electronic
 superconductors mentioned above \cite{Sarma}.
We shall show that the next phase of the CFL phase down in the density
 is not necessarily the gCFL phase, but strongly depends on the
 coupling constant in the scalar diquark channel even at zero
 temperature: 
The gCFL phase is found to appear only in the small coupling
 regime; this fragileness of the gCFL phase with the dynamical quark
 condensate will be shown to be understandable in a rather
 model-independent way.
We shall show that the most favorable phases realized in a wide
 parameter window are the g2SC, 2SC, and unpaired neutral phases.

We start from the following extended three flavor Nambu--Jona-Lasinio
 (NJL) Lagrangian density with the diquark coupling $G_d$, and the
 scalar coupling $G_s$ \cite{HS02,SRP02}.
\bea
\label{lag}
{\mathcal L}&=& \bar{q}(i\feyn{\p} - \bfm{m}_0 + \bfm{\mu}\gamma_0)q%
     + \frac{G_d}{16}\sum_{\eta=1}^{3}%
       \big[(\bar{q}P_\eta^t\bar{q})(^tq\bar{P}_\eta q)\big]\nn
 & & + \frac{G_s}{8N_c}%
     \big[(\bar{q}\bfm{\la}_Fq)^2+ (\bar{q}i\ga_5\bfm{\la}_F q)^2\big].
\eea
Here, $\bfm{\la}_F=\{\sqrt{2/3}\bfm{1},\vec{\la}_F\}$ are the unit
 matrix and the Gell-Mann matrices in the flavor space. $P_\eta$ is
 defined as in Ref.~\cite{AKR04b}
\begin{equation}
  (P_\eta)^{ab}_{ij}=i\gamma_5C\epsilon^{\eta ab}\epsilon_{\eta ij}%
  \quad\mbox{no sum over index $\eta$}
\end{equation}
and $\bar{P}_\eta=\gamma_0 (P_\eta)^\dagger\gamma_0$. 
$a,b,\cdots$ and $i,j,\cdots$ represent the color and flavor indices,
 respectively. 
The second term in \Eqn{lag} simulates the attractive interaction
 in the color anti-triplet, the flavor anti-triplet and in the $J^P=0^+$
 channel in QCD. 
$\bfm{m}_0={\rm diag.}\{m_u,m_d,m_s\}$ is the current-quark mass matrix;
  the full lattice QCD simulation shows that
 $m_{u,d}(2\GeV)=3\,$--$\,4\MeV$ and $m_s(2\GeV)=80\,$--$\,100\MeV$
 \cite{cp-pacs}.
We take the chiral SU(2) limit for the u, d sector ($m_u=m_d=0$)
 and $m_s=80\MeV$ throughout this paper. 
These values might slightly underestimate the effect of the current
 masses.

In order to impose the color and electric neutrality, we introduce the
  chemical potential matrix $\bfm{\mu}$ in the color-flavor space as
\begin{equation}
 \bfm{\mu}^{ab}_{ij}=\mu \de_{ab}B_{ij} -\mue \de^{ab}Q_{ij}%
 + \mu_3 \de_{ij}T_3^{ab} + \mu_8 \de_{ij}T_8^{ab}.
\end{equation}
$B_{ij}=\de_{ij}$ and $Q_{ij}={\rm diag.}\{2/3,-1/3,-1/3\}$ count
  baryon number and electric charge of quarks, respectively.
$T_3^{ab}={\rm   diag.}\{1/2,-1/2,0\}$ and
  $T_8^{ab}={\rm}\{1/3,1/3,-2/3\}$ are the diagonal generators of the
  color SU(3).
In the numerical analysis, we shall adopt the three-momentum cutoff
  $\La=800\MeV$ and the scalar coupling constant $G_s$ giving the
  vacuum constituent quark mass $400\MeV$ in the chiral limit, for
  comparison with the results in Refs.~\cite{AKR04a,AKR04b,FKR04};
these parameter values give larger condensates than those used in
  \cite{HK94} and \cite{HS02,SRP02,buballa_hb}.

We treat the diquark coupling constant $G_d$ as a simple parameter,
  although the perturbative one-gluon exchange vertex ${\cal   L}_{\rm
  int}=-(g^2/2)\bar{q}\gamma_\mu(\la_a/2)q\bar{q}\gamma^\mu(\la_a/2)q$, 
  which is valid at extremely high density, tells us that $G_d/G_s=1/2$
  with $N_c=3$ \cite{SRP02,eta01}.
Furthermore, we shall use, instead of $G_d$, the gap energy ($\De_0$) in
  the pure CFL phase at $\mu=500\MeV$ and $T=0$ in the chiral SU(3)
  limit, as was done in Refs.~\cite{AKR04a,AKR04b,FKR04}.

We evaluate the thermodynamic potential in the mean-field approximation;
\bea
  \Om&=&\frac{4}{G_d}\sum_{\eta=1}^{3}\De_\eta^2+%
  \frac{N_c}{G_s}\sum_{i=1}^{3}(M_i-m_i)^2 \nn
  &&-\frac{T}{2}\int\frac{d\bfm{p}}{(2\pi)^3}\Tr{\rm Log}%
    \left[S^{-1}(i\omega_n,\bfm{p})\right]+\Om_e,\label{det}
\eea
where 
\bea
  \De_\eta&=&\frac{G_d}{8}\lt ^tqP_\eta q\gt,\\
  \bfm{M}&=&\left( \begin{array}{ccc}
  M& 0 & 0 \\
  0& M & 0 \\
  0& 0 & \Ms
  \end{array}\right)\nn
  &=&\bfm{m}_0+\frac{G_s}{2}%
  \left( \begin{array}{ccc}
  \lt\bar{u}u\gt& 0 & 0  \\
  0 & \lt\bar{d}d\gt& 0 \\
  0 & 0 &\lt\bar{s}s\gt 
  \end{array}\right),\label{model}
\eea
are the gap parameter matrix and the constituent quark mass matrix,
  respectively, and $S$ denotes the $72\times 72$ Nambu-Gor'kov
  propagator  defined by
\begin{equation}
   S^{-1}(i\omega_n,p)=\left( \begin{array}{cc}
  \feyn{p}+{\bfm{\mu}}\gamma_0-\bfm{M}& \sum_\eta
   P_\eta\De_\eta  \\
  \sum_\eta \bar{P}_\eta\De_\eta & 
  ^t\feyn{p}-{\bfm{\mu}}\gamma_0+\bfm{M}
  \end{array}\right),
\end{equation}
   with $\feyn{p}=i\omega_n\gamma_0-\bfm{p}\cdot\bfm{\gamma}$.
Finally, $\Om_e$ is  the contribution from the massless electrons
\bea
  \Om_e&=&-\frac{\mue^4}{12\pi^2}\nn
  &&-2T\int\frac{d\bfm{p}}{(2\pi)^3}\sum_{\xi=\pm}%
  \left[\log(1+e^{-|\mue-\xi p|/T})\right].
\eea
The functional determinant in \Eqn{det} can be evaluated using the
   method given in the literature \cite{AKR04b,FKR04}.
The optimal values of the variational parameters $\De_\eta$, $M$ and $\Ms$
   must satisfy the stationary condition (the gap equations);
\begin{equation}
  \frac{\p\Om}{\p\De_\eta}=0,\,\,\frac{\p\Om}{\p M}=0%
  \,\,\mbox{and}\,\,\frac{\p\Om}{\p \Ms}=0.\label{eq:gapeq}
\end{equation}
In order to clarify the effects of the chiral condensation on the
   diquark pairing, we shall also reexamine the case in which
   the dynamical chiral condensates are not incorporated
   \cite{AKR04a,AKR04b,RSR04,FKR04};
\bea
 \frac{\p\Om}{\p\De_\eta}\biggl|_{\bfsm{m_0}=\bfsm{M}}=0.\label{eq:gapeq0}
\eea
Here the quark mass $\Ms$ is regarded as the in-medium strange 
   quark mass and will be varied by hand.
Our task is to search the minimum of the effective potential through
   solving these  gap equations under the local electric and color
   charge neutrality conditions;
\begin{equation}
  \frac{\p\Om}{\p\mue}=0,\,\,\frac{\p\Om}{\p\mu_3}=0%
  \,\,\mbox{and}\,\,\frac{\p\Om}{\p\mu_8}=0.
\end{equation}

\begin{table*}[t]
 \begin{tabular}{|r||ccccc|c|cccc|}
  \hline
  \multicolumn{1}{|c||}{\bf Name of}& \multicolumn{5}{c|}{\bf Gap and
  mass parameters} & {\bf Conditions} & \multicolumn{4}{c|}{\bf Gapless
  quarks} \\
  \multicolumn{1}{|c||}{\bf Phase}& $\De_1(ds)$ & $\De_2(us)$ &
  $\De_3(ud)$ & $M$ & $\Ms$ & {\bf for chemical potentials}
  &~~($ru$-$gd$-$bs$)~~&~~($bd$-$gs$)~~&~~($bu$-$rs$)~~&~~($gu$-$rd$)~~\\
  \hline
  CFL (9)
  & $\De_1$ & $\De_2$ & $\De_3$ &
  & $\Ms$ &[$\mue=0$] &\multicolumn{4}{c|}{all quark modes are fully
  gapped}\\ \hline
  gCFL (7)
  & $\De_1$ & $\De_2$ & $\De_3$
  & & $\Ms$ & $\de\mu_{bd\mbox{-}gs(bu\mbox{-}rs)}%
      +\frac{\Ms^2}{4\mu}\agt\De_{1(2)}$ & & $bd$ & $bu$  & \\ \hline
  uSC (6)
  & & $\De_2$ & $\De_3$ & & $\Ms$
  & [$\mue=0$] & $gd$-$bs$ (1) &($bd$,~$gs$) & & \\ \hline
  guSC (5)
  & & $\De_2$ & $\De_3$ & &
  $\Ms$ & $\de\mu_{bu\mbox{-}rs}+\frac{\Ms^2}{4\mu}\agt\De_2$  &
  $gd$-$bs$ (1) &  ($bd$,~$gs$) & $bu$ & \\ \hline\hline
  2SC (4)
  & & & $\De_3$ & & $\Ms$ &
  [$\mu_3=0] $ & $bs$ & ($bd$,~$gs$) & ($bu$,~$rs$) & \\ \hline
  g2SC (2)
  & & & $\De_3$ & & $\Ms$ &
  [$\mu_3=0$], $\de\mu_{gd\mbox{-}ru}=\de\mu_{rd\mbox{-}gu}>\De_3$ &
  $gd$,~$bs$ & ($bd$,~$gs$) & ($bu$,~$rs$) & $rd$\\ \hline
  dSC (6)
  & $\De_1$ & & $\De_3$ & & $\Ms$
  & & $ru$-$bs$ (1) & & ($bu$,~$rs$) & \\ \hline
  gdSC (5)
  & $\De_1$ & & $\De_3$ & &
  $\Ms$ & $\de\mu_{bd\mbox{-}gs}+\frac{\Ms^2}{4\mu}\agt\De_1$ &
  $ru$-$bs$ (1)& $bd$ & ($bu$,~$rs$) & \\ \hline\hline
  2SCus (4)
  & & $\De_2$ & & & $\Ms$ & &
  $gd$  & ($bd$,~$gs$)&  & ($bu$,~$rs$) \\ \hline\hline
  UQM (0)
  & & & & & $\Ms$ &
  [$\mu_3=\mu_8=0]$ &\multicolumn{4}{c|}{all quarks are ungapped.} \\
  \hline
  $\chi$SB (0)
  & & & &$M$ &$\Ms$ &
  [$\mu_3=\mu_8=0]$ & \multicolumn{4}{c|}{all quarks are massive.} \\ 
  \hline
 \end{tabular}
 \caption[]{%
 The nonzero gap parameters, some conditions between gaps and chemical
   potentials, and the gapless quarks in each phase. 
 We abbreviate the unpaired neutral quark matter with nonzero
   $\lt\bar{s}s\gt$ to ``UQM'', and the chiral-symmetry broken phase as
   ``$\chi$SB'', respectively. 
 (g) means the gapless version of the pairing state.
 The number in the parenthesis in the first column represents the number
   of gapped quasi-quark modes. ``$gd$-$bs$ (1)'' means that one of the
   linear combination of $gd$ and $bs$ quarks remains gapless. 
   $\de\mu_{gd\mbox{-}bs}$ denotes the chemical potential difference
   $\equiv(\mu_{gd}-\mu_{bs})/2$, and so forth.
 The equations for chemical potentials which are necessarily satisfied
   in the given phase are written in a bracket in the third column.
 }
 \label{gaps}
\end{table*}
Before going into a numerical computation, one must specify the phase
   characterized by the various patterns of the mean fields and the
   chemical potentials ($\mue, \mu_3, \mu_8$); 
comparing the value of $\Om$ obtained for each phase, one can
   determine which phase is realized for given $\mu$ and $\De_0$.
In the present analysis, we consider the states listed in
   TABLE~\ref{gaps} as  possible phases to be realized. 
These phases are described in the following sub-model spaces defined by
   the parameter sets in the parenthesis,
respectively; within which the gap equation for
   each phase is solved.

\vspace*{0.1cm}
\noindent
{\bf Set.~1}~ $(\mue,M,\Ms)$: 
This parameter space can model the $\chi$SB phase 
  and the neutral unpaired quark matter (UQM).
  The dynamical condensates in all flavor sectors
  $(\lt\bar{u}u\gt,\lt\bar{d}d\gt,\lt\bar{s}s\gt)$ are 
  accompanied in the $\chi$SB phase, while in the UQM phase,
  only $\lt\bar{s}s\gt$ is non-vanishing.\\
{\bf Set.~2}~ $(\De_3,\mue,\mu_8,\Ms)$: 
In this parameter space, the (g)2SC and UQM phases are described.\\
{\bf Set.~3}~ $(\De_2,\mue,\mu_3,\mu_8,\Ms)$:
The 2SCus and UQM phases are described in this parameter space.\\
{\bf Set.~4}~ $(\De_2,\De_3,\mue,\mu_3,\mu_8,\Ms)$: 
The (g)uSC, (g)2SC and UQM phases are described in this space.\\
{\bf Set.~5}~ $(\De_1,\De_3,\mue,\mu_3,\mu_8,\Ms)$: 
The (g)dSC, (g)2SC and UQM phases are described in this space
\cite{Mat04}. \\
{\bf Set.~6}~ $(\De_1,\De_2,\De_3,\mue,\mu_3,\mu_8,\Ms)$: 
The (g)CFL, (g)uSC and UQM phases are described in this space.

\vspace*{0.1cm}
\noindent
We numerically confirmed that other phases such as 2SCds or sSC phases,
  which are described with the parameter set
  $(\De_1,\mue,\mu_3,\mu_8)$ or $(\De_1,\De_2,\mue,\mu_3,\mu_8)$,
  respectively, never get realized as the ground state at $T=0$.
In this Letter, we restrict ourselves to the zero temperature case,
  leaving an analysis on the  $T\ne0$ case for a future publication.

\begin{figure*}[t]
 \begin{minipage}{0.49\textwidth}
  \includegraphics[width=0.8\textwidth,clip]{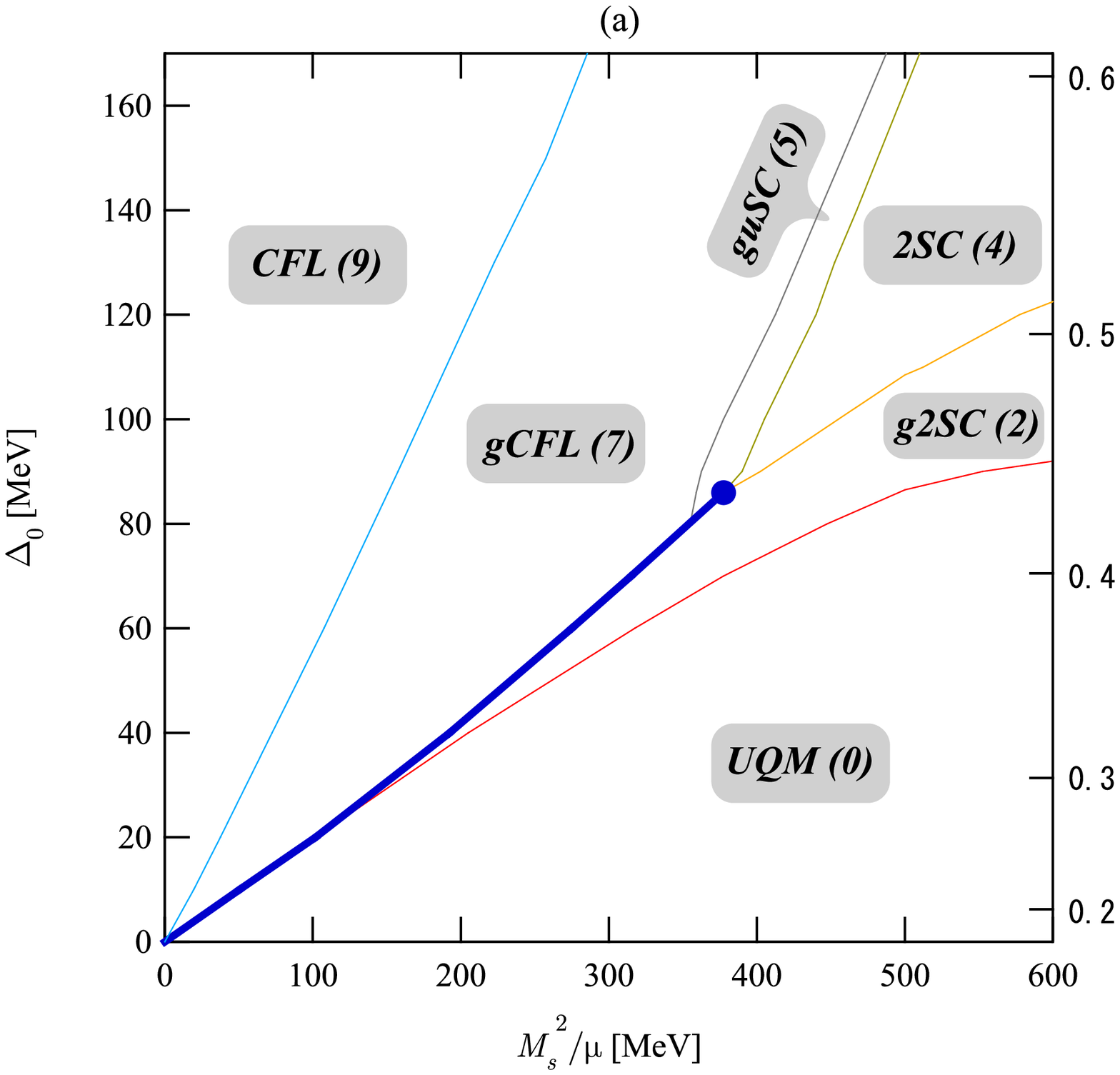}
 \end{minipage}%
 \hfill%
 \begin{minipage}{0.49\textwidth}
  \includegraphics[width=0.8\textwidth,clip]{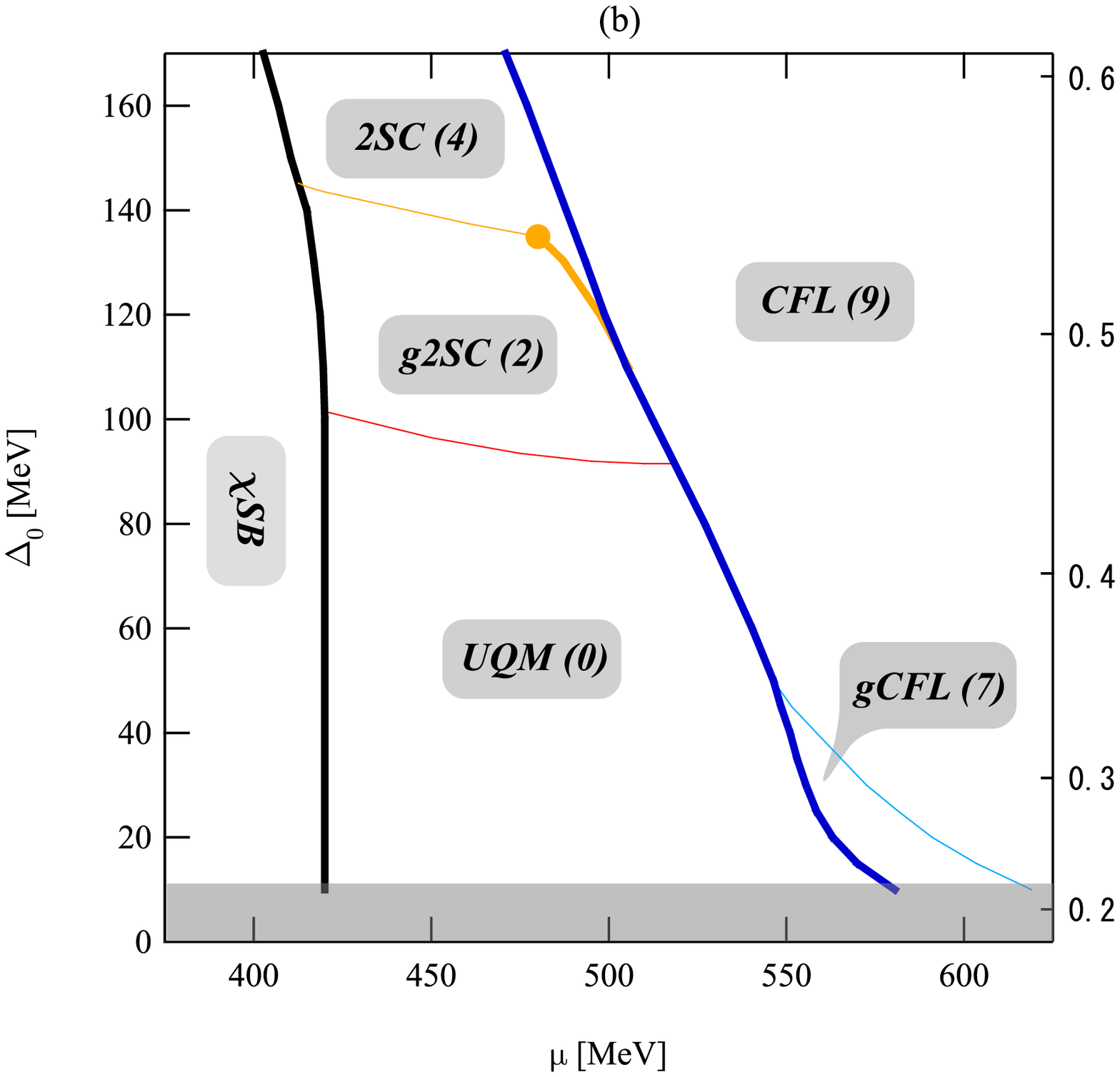}
 \end{minipage}
 \caption[]{%
(a)~Phase diagram in $(\De_0,\Ms^2/\mu)$ plane:~%
The gap equations \Eqn{eq:gapeq0} under the neutrality
  constraints are solved with varying $\Ms=m_s$ by hand, while the
  quark chemical potential is fixed at $\mu=500\MeV$ as in
  Refs.~\cite{AKR04a,AKR04b,RSR04,FKR04}.
The scale of the  vertical axis on the right-hand side represents
  the value of $\eta=G_d/G_s$ for the corresponding diquark coupling
  $\De_0$. 
The number in a parenthesis attached to each phase name in the figure
  denotes the number of gapped quasi-quarks. 
(b)~Phase diagram in $(\De_0,\mu)$ plane:~%
Continuous (first order) transition lines are represented by thin (bold)
  lines. 
 Although we did not determine the phase boundaries in the
 dark-shaded area where a rather better precision is required for the
 comparison of the potentials, we confirmed that any new phase structure
 does not appear in this region.
 }
 \label{fig:00}
\end{figure*}

We first present the phase diagram {\em without} the dynamical chiral
  condensates with $\Ms$ being varied by hand for a fixed quark chemical
  potential $\mu=500\MeV$; 
the ground state is searched with the aid of the gap equation
  \Eqn{eq:gapeq0} which gives a candidate of the ground state.
FIG.~\ref{fig:00}(a) shows an entire phase diagram in the
  $(\Ms^2/\mu,\De_0)$ plane; we notice that this phase diagram is
  consistent with the one given in \cite{FKR04}, in which the phase
  structure only for several values of $\De_0$ is given.
One may notice the following points:\\
(1) The gCFL phase always exists as the next phase of the CFL phase down
  in density, irrespective of the value of the diquark coupling $\De_0$.
In addition, the parameter region of $\Ms^2/\mu$ for accommodating the
  gCFL phase grows as the diquark coupling constant increases.\\
(2) The stronger the coupling $\De_0$, the richer the phase structure:
  The CFL phase turns into the UQM phase through successive
  transitions; gCFL $\to$ guSC $\to$ 2SC $\to$ g2SC. 
Accordingly, the number of frozen degrees of freedom (gapped quarks)
  decreases as $9\to7\to5\to4\to2\to0$ towards lower density.

Now let us examine how the above features are changed once the strange
  quark mass is determined dynamically through \Eqn{eq:gapeq}.
The resulting phase diagram in the ($\mu,\De_0$) plane is shown in
  FIG.~\ref{fig:00}(b); 
$\Ms$ is now determined dynamically and thus becomes a function of $\mu$
  and $\De_0$. 
The following should be notable from the figure:\\
(1) Although the CFL phase is present still in all the diquark coupling
  region, the chemical potential window for realizing the gCFL phase
  shrinks with the increasing coupling constant, and eventually
  closes at $\De_0\sim 50\MeV$.
  This is highly in contrast with the case described above.\\
(2) As the coupling is increased, the following phases appear
  successively as the next phase of the CFL phase down in density;
  the gCFL phase, the UQM phase, the g2SC phase, and finally the 2SC
  phase.

We can clearly divide the entire phase diagram into four distinct
  regions according to which phase comes in as the next phase  down from
  the CFL phase.
(i) The {\em weak coupling regime} ($\De_0\alt50\MeV$); 
  the gCFL phase exists between the UQM and CFL phases.
(ii) The {\em moderate coupling regime} ($50\MeV\alt\De_0\alt90\MeV$);
  the gCFL phase ceases to be the secondly densest phase, and is taken
  over by the UQM phase, which is nearly two-flavor quark matter with
  large $\lt\bar{s}s\gt$ condensate.
  We stress that the superconductivity itself is destroyed before the
  gCFL phase sets in when the density is decreased.
(iii) The {\em strong coupling regime} ($90\MeV\alt\De_0\alt140\MeV$); 
  a gapless superconducting phase reappear but only with the two flavors
  being involved, which is called the g2SC phase. 
  Also a large chiral condensate $\lt\bar{s}s\gt$ is accompanied in this
  phase. 
(iv) The {\em extremely strong coupling regime} ($140\MeV\alt\De_0$);
  any gapless superconductivity vanishes and the fully-gapped 2SC phase
  is realized. 
  It is worth mentioning that the g2SC $\to$ 2SC transition with the
  increasing diquark coupling near the CFL phase seen in
  Fig.~\ref{fig:00}(b) is not of a second order but of a first order
  with a small jump in the strange quark mass and density. 
  The first order phase transition ends at the point denoted by a large
  dot in FIG.~\ref{fig:00}(b). 
  For smaller chemical potential than that at this point, the
  2SC $\to$ g2SC transition is continuous and essentially the same as that
  found in the original paper \cite{HS02}, because the on-shell strange
  quarks are not present in this region due to a large dynamical mass of
  the strange quark.

\begin{figure*}[tp]
 \begin{minipage}{0.49\textwidth}
  \includegraphics[width=0.8\textwidth,clip]{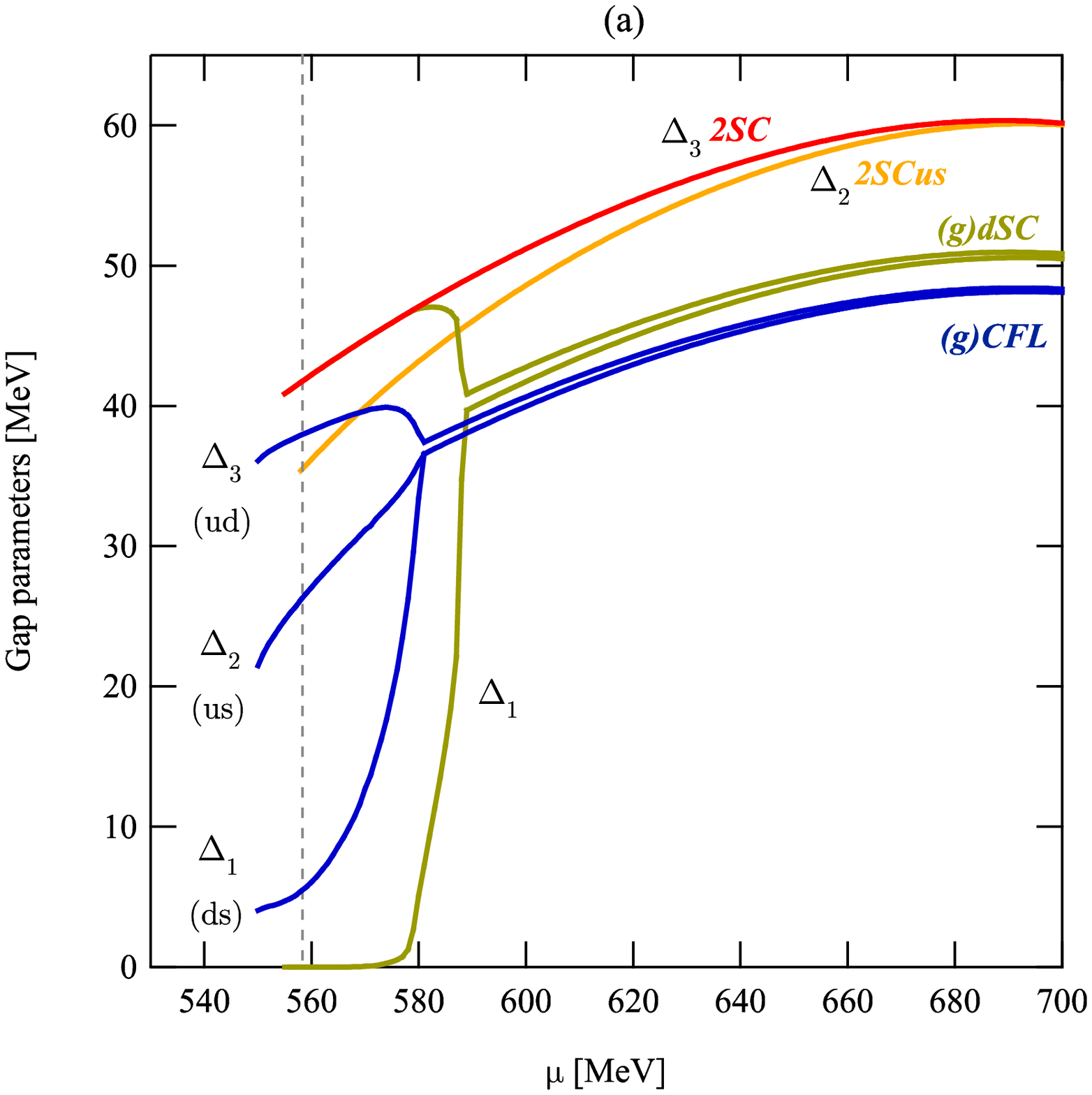}
 \end{minipage}%
 \hfill%
 \begin{minipage}{0.49\textwidth}
  \includegraphics[width=0.8\textwidth,clip]{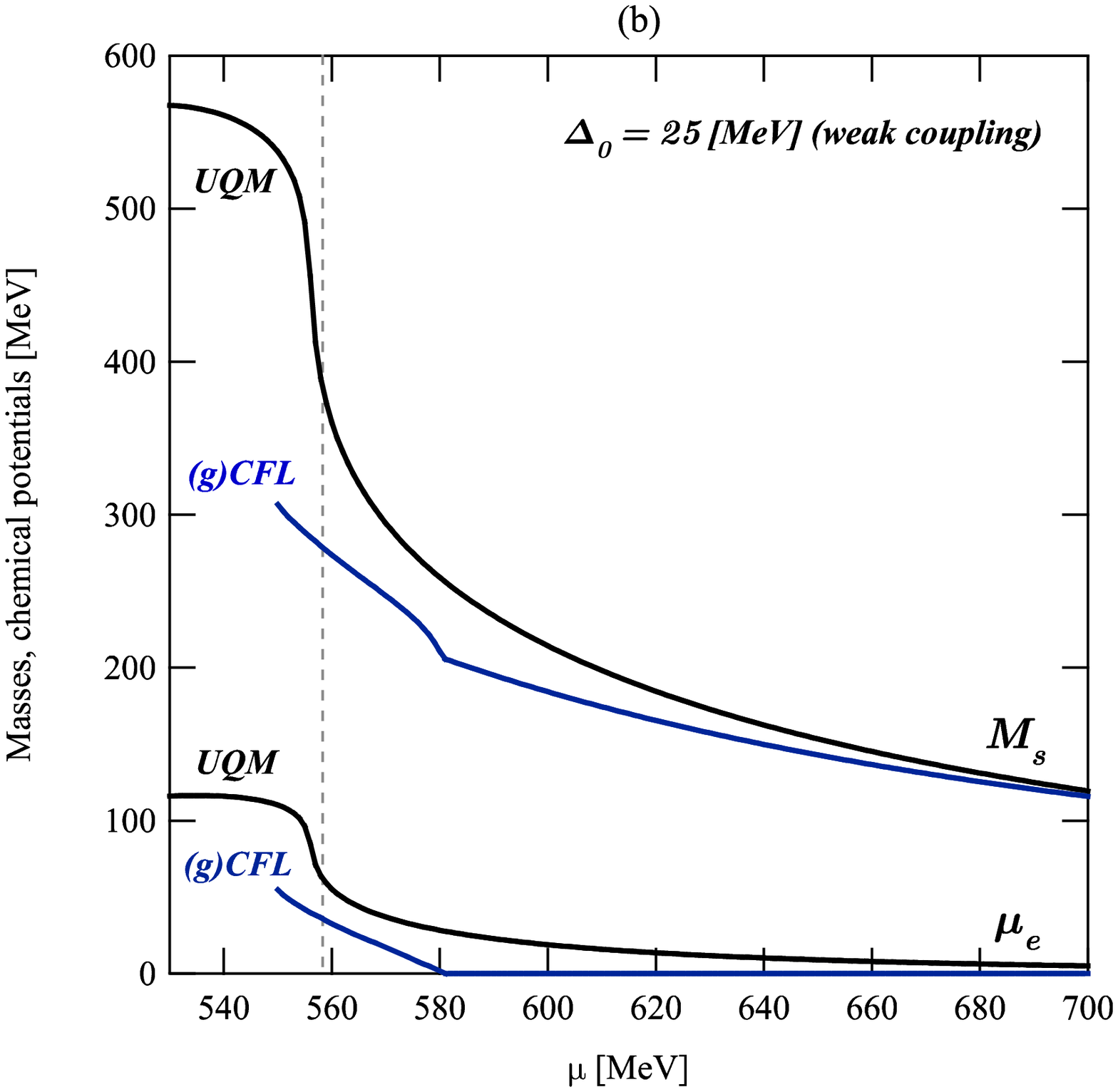}
 \end{minipage}
 \caption[]{%
 (a)~The gap parameters for each state at a weak coupling
   $\De_0=25\MeV$. 
   We remark that the states other than the (g)CFL and UQM phases are  
   realized only as metastable states for all the region of $\mu$;
   nevertheless the 2SC, 2SCus and (g)dSC phases give the global 
   minimum in the corresponding sub-model spaces, namely, Set.~2,
   Set.~3, and Set.~5, respectively.
   The vertical dashed line represents the point $\mu\cong558\MeV$ where
   the UQM phase takes over the gCFL phase, blow which the UQM phase
   with massive strange quarks accordingly becomes the ground state of
   the system. 
   The CFL phase turns into the gapless CFL phase at the critical chemical
   potential $\mu=\mu_*=581.1\MeV$.
 (b)~The state-dependent strange quark masses and the electron chemical
   potentials for the (g)CFL and UQM phases at the same coupling as
   in (a).
 }
 \label{fig:01}
\end{figure*}

So far the global structure of the phase diagram.
Next let us discuss in detail how the gCFL phase disappears for the
  stronger diquark coupling, examining closely two typical coupling cases;
  $\De_0=25\MeV$ ({\em weak coupling}), $\De_0=60\MeV$ ({\em moderate
  coupling}).
We shall also show the $\mu$ dependence of gaps in the phases which are
  actually not realized as the ground state, for completeness.

FIG.~\ref{fig:01}(a) shows the gap parameters as functions of $\mu$ for
  several phases.
At high chemical potentials $\mu\agt580\MeV$, the ground state is 
  in the CFL phase.
As the density is decreased, the stress energy
$M_s^2(\mu)/\mu$
increases in the symmetric CFL
  pairing; notice that as the density is decreased, the in-medium
  strange quark mass $M_s(\mu)$ increases, which causes a further
  increase of the stress $M_s^2(\mu)/\mu$.
Accordingly, the phase suffers from a slight distortion
  in the gaps ($\De_1=\De_2\alt\De_3$); 
  nevertheless the CFL phase persists down to the critical chemical
  potential $\mu=\mu_*\cong581.1\MeV$, at which the effective chemical
  potential difference in ($bd$-$gs$) sector, $\de\mu_{\rm
  eff}(bd\mbox{-}gs)\equiv\de\mu_{bd\mbox{-}gs}+\frac{\Ms^2}{4\mu}$,
  reaches almost the magnitude of the gap $\De_1=\De_2$.
This transition is essentially the $\Qtilde$-insulator-to-metal
  transition discussed in \cite{AKR04a,AKR04b}, with $\Qtilde$ being the
  unbroken U(1) charge in the CFL phase \cite{reviews,ARW98}.
We notice that the onset condition of this metal-insulator transition,
  $\Ms^2(\mu_*)/2\mu_*\cong \De_1(\mu_*)$ \cite{AKR04a}, works well even
  when the strange quark mass is treated as a dynamical variable.
As the density is decreased further, the UQM phase with massive
  strange quarks finally takes over the gCFL phase at $\mu\cong558\MeV$,
  which is denoted by the dashed line in FIG.~\ref{fig:01}.
This unlocking transition gCFL $\to$ UQM is a first order as was found in
  Refs.~\cite{AKR04a,AKR04b,FKR04}, although the transition point is
  somewhat shifted to a lower $\Ms(\mu)$ (higher $\mu$) than is obtained
  in these papers.

FIG.~\ref{fig:01}(b) shows the state-dependent strange quark mass as a
  function of $\mu$ in the (g)CFL and the UQM phases.  
One should notice that the generation of the dynamical strange quark
  mass is suppressed in the (g)CFL phase in comparison with that in the
  UQM phase.
This is because the (g)CFL paring requires a Fermi-momentum matching
  among all the species as much as possible, which is achieved by
  suppressing the dynamical generation of the strange quark mass;
it is also to be noticed that a better Fermi-momentum matching also
  lowers the energy cost for imposing charge neutrality.
On the other hand, the UQM phase does not need such a Fermi-momentum
  matching, and thus can gain the condensation energy in the chiral
  $\lt\bar{s}s\gt$ sector.
In other words, the first-order unlocking gCFL $\to$ UQM phase
  transition is brought about by the competition between the following
  two factors; 
(i) reducing the neutrality costs by matching the Fermi momenta
  of three species in the gCFL phase, and
(ii) increasing the condensation energy gain in the chiral
  $\lt\bar{s}s\gt$ sector in the UQM phase.
The former effect (i) is underestimated in the previous work
   \cite{AKR04a,AKR04b,FKR04} because the strange quark mass is treated
   as a simple parameter, \ie, $\Ms({\rm UQM})=\Ms({\rm CFL})$.
The latter effect (ii) is taken into account for the first time in the
  present work.
As a consequence, the transition density (strange quark mass) is
  somewhat larger (smaller) in comparison with the results in
  \cite{AKR04a,AKR04b,FKR04}; 
in fact the transition takes place at $\Ms^2/\mu\sim 4\De_1(\mu_*)$ in
  the present work, while $\Ms^2/\mu\sim 5\De_1(\mu_*)$ in
  Refs.~\cite{AKR04a,AKR04b,FKR04}, with $\De_1(\mu_*)$ being the gap in
  the ($bd$-$gs$) sector at the onset point of the gCFL phase.
We emphasize that this tendency of destabilizing the (g)CFL phase by the
  dynamical generation of chiral condensate should hold generically, not
  depending on a model used.

\begin{figure}[t]
  \includegraphics[width=0.4\textwidth,clip]{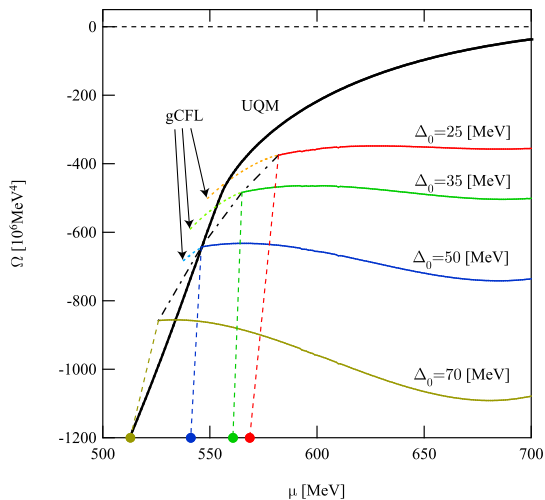}
 \caption[]{The thermodynamic potentials $\Omega$ for the UQM 
 phase and the CFL phases with various couplings as functions of $\mu$.
 We have chosen the zero of the potential such that the unpaired
 neutral quark matter with bare ($M_s=80\MeV$) strange quarks has
 $\Omega=0$.
 The CFL phases turn into the gapless phases at the points where the
 bold lines change into the dashed lines.
 These gCFL-CFL transition points are linked by the dash-dotted line.
 In the horizontal axis, we also put large dots, which denote the values
 $M_s^2(\mu_*)/2\Delta_1(\mu_*)$ for corresponding $\Delta_0$.
 }
 \label{fig:02}
\end{figure}

How do the above features change when the diquark coupling is raised?
One might naively expect that the window in
$\mu$ for the gCFL realization
becomes wide, thinking that the gCFL phase should become more robust in
the stronger coupling regime.
It is, however, not the case; in fact, we have no longer a window for
the gCFL phase in the moderate coupling, for instance, at
$\De_0=60\MeV$. (See FIG.~\ref{fig:00}(b).)
This disappearance of the gCFL phase in the stronger coupling can be
  nicely understood with the aid of the thermodynamic potential.
In FIG.~\ref{fig:02}, we show the thermodynamic potentials for the UQM
  phase and the CFL phases with $\Delta_0=25,\,35,\,50,\,\mbox{and
  }70\MeV$.
We first notice that $\mu=M_s^2(\mu_*)/2\Delta_1(\mu_*)$ denoted by
  large dots on the horizontal axis gives a quite
  good guide for the critical chemical potential $\mu_*$ for the
  CFL-gCFL transition.
Accordingly, the transition point shifts to a lower $\mu$ as the diquark
  coupling is increased and eventually moves to the left of the UQM line
  which rapidly falls for $\mu\alt550\MeV$  because of the quite large
  chiral condensate realized in the UQM phase as is clearly seen in
  FIG.~\ref{fig:01}(b).
In particular, for $\Delta_0=70\MeV$, the CFL phase becomes metastable
  against the UQM phase before the gCFL phase sets in.
In other words, the UQM phase takes over the CFL phase before the
  gapless dispersions of $bd$ and $bu$ quarks get to be realized.
In short, although the CFL phase becomes more stable with the
 larger diquark coupling, making the gCFL-CFL transition
  point $\mu_*$ lower, the UQM phase overwhelms all the paired phases
  because of the large $\lt\bar{s}s\gt$-condensation energy.
We also remark that at the end points of the (g)CFL lines in
  FIG.~\ref{fig:02}, the (g)CFL states cease to exist even as a
  metastable state, nor as a local maximum like the unstable Sarma state
  \cite{Sarma}.

Finally let us see the reason why the (g)2SC phase comes as the next
  phase down in density as the diquark coupling  is increased further.
As the coupling is increased, the UQM phase is expected to
  experience successive second order phase transitions firstly to the
  g2SC phase and then to the 2SC phase if the strange quarks
  are not present in the system \cite{HS02}.
Our full calculation shows that the on-shell strange quarks are absent
  in the system at $\mu\alt450\MeV$ because of a large $\Ms$;
thus the series of transitions UQM $\to$ g2SC $\to$ 2SC becomes
  essentially the same as in \cite{HS02}.
At relatively higher chemical potential, however, the small amount of
  the strange quarks are present, which brings about a non-trivial
  tricritical point on the g2SC $\to$ 2SC transition line in the phase
  diagram (see the large dot in FIG.~\ref{fig:00}(b)).
At higher $\mu$ than this point, the g2SC $\to$ 2SC transition is a first
  order with jumps in physical quantities as mentioned before.
We notice that this first order transition is also caused by the
  competition between the chiral $\lt\bar{s}s\gt$ condensation in the g2SC
  phase and the pairing energy gain in the fully gapped 2SC order.
In fact, the transition is from a 2SC phase (2SC+s) with a small strange
  quark density to a nearly two-flavor g2SC phase with larger vacuum
  $\lt\bar{s}s\gt$ condensate along the first order line.
The 2SC phase tends to lower the density mismatch between $n_u$ and
  $n_d$, and thus needs more strange quarks as well as electrons for the
  electric neutrality than in the g2SC phase.
As a result, the 2SC+s phase tends to reduce the dynamical strange quark
  mass.
In contrast, the g2SC phase has extra d quarks accumulated in the
  momentum shell, thus requires less strange quarks.
For this reason, the g2SC phase can have a larger dynamical chiral
  condensate $\lt\bar{s}s\gt$.

In this Letter, we have made an  analysis on the interplay between the
  chiral and diquark condensations in the three-flavor neutral quark
  matter using an extended NJL model.
We have shown that which phase appears next to the CFL phase strongly
  depends on the strength of the diquark coupling;
  as the diquark coupling is increased, the gCFL, the UQM, the g2SC and
  finally the 2SC phase appear as the next phase down in density.
The disappearance of the gCFL phase in the strong coupling regime and
  the emergence of a non-trivial tricritical point on the 2SC $\to$ g2SC
  transition line are qualitatively understandable in terms of the
  competition among the chiral-condensation energy, the gain through the
  pairing and the energy cost due to neutrality constraints.
Although the present calculation is performed with a
  specific value of the cutoff $\Lambda=800\MeV$, we have confirmed
  that our central result, namely, the shrinkage of the gCFL window 
 with the increasing diquark coupling,
 is unaffected with the change of $\Lambda$ in the
  range $600\sim1000\MeV$.
In a longer paper, we shall present a more detailed analysis on 
  the nature of phase transitions obtained here, giving some physical
  quantities including strange and isospin densities in each phase.

In the present work, we have restricted ourselves to the case with
  vanishing temperature.
It would be interesting to study the competition between the chiral and
  diquark condensations at finite temperature, and to examine the
  robustness or fragileness of the phases obtained here.
The extension of this work to the nonzero temperature is straightforward
  and will be discussed elsewhere.
Finally, we have not considered here the possibility of the quantum
  inhomogeneous state \cite{Crystal,hetro} and the possible meson
  condensation in the CFL phase \cite{kaoncondensation} from the
  beginning.
Also we did not take care of a potential instability due 
to the
  imaginary Meissner mass in the gapless phases \cite{Unstable}.
A detailed analysis including all these possibilities is certainly a
  challenge but needed for our deeper understanding of the QCD phase
  diagram.

\vspace*{0.1cm}
\noindent
The authors are grateful to M.~Alford for his interest in 
this work and valuable comments which led to FIG.~\ref{fig:02}.
One of the authors, H. A. thanks M. Asakawa for encouragement.
H. A. is supported by the Fellowship program, Grant-in-Aid for the
   21COE, ``Center for Diversity and Universality in Physics'' 
   at Kyoto University.
M. K. is supported by Japan Society for the Promotion of Science for
   Young Scientists.
T. K. is supported by Grant-in-Aide for Scientific Research by
   Monbu-Kagaku-sho (No.\ 14540263).
This work is supported in part by a Grant-in-Aid for the 21st Century 
   COE ``Center for Diversity and Universality in Physics''.

\end{document}